\documentstyle[prd,aps]{revtex}
\begin{document}
\input epsf \renewcommand{\topfraction}{0.8}

\catcode`@=11
\def\gsim{\mathrel{\mathpalette\@versim>}}
\def\@versim#1#2{\lower0.2ex\vbox{\baselineskip\z@skip\lineskip\z@skip
      \lineskiplimit\z@\ialign{$\m@th#1\hfil##$\crcr#2\crcr\sim\crcr}}}
\catcode`@=12
\catcode`@=11
\def\lsim{\mathrel{\mathpalette\@versim<}}
\def\@versim#1#2{\lower0.2ex\vbox{\baselineskip\z@skip\lineskip\z@skip
       \lineskiplimit\z@\ialign{$\m@th#1\hfil##$\crcr#2\crcr\sim\crcr}}}
\catcode`@=12
\draft
\twocolumn[\hsize\textwidth\columnwidth\hsize\csname
@twocolumnfalse\endcsname

\title{Extended Inflation with an Exponential Potential}
\author{Mikel Susperregi$^{1}$ and Anupam Mazumdar$^{2}$}
\address{$^{1}$ Astronomy Unit, School of Mathematical Sciences, 
Queen Mary \& Westfield College,\\ University of London,    
London E1 4NS, United Kingdom\\
$^{2}$ Astronomy Centre, University of Sussex, Falmer, 
Brighton BN1 9QJ, United Kingdom}
\date{\today}
\maketitle
\begin{abstract}
In this paper we investigate extended inflation with 
an exponential potential $V(\sigma)= V_0\,e^{-\kappa\sigma}$, which 
provides a simple cosmological scenario where the distribution of 
the constants of Nature is mostly determined by $\kappa$. 
In particular, we show that this theory predicts 
a uniform distribution for the Planck mass at the end of inflation, 
for the entire ensemble of universes that undergo stochastic
inflation. Eternal inflation takes place in this scenario for 
a broad family of initial conditions, all of which lead up to 
the same value of the Planck mass at the end of inflation. 
The predicted value of the Planck mass is consistent with the 
observed value within a comfortable range of values of the 
parameters involved.
\end{abstract}

\vskip2pc]


\section {Introduction}

The idea of a varying $G$ based on anthropic arguments, or what 
would eventually become concisely enunciated anthropic arguments,
dates back to Dirac \cite{dirac} and later Sciama \cite{sciama}. 
Their ideas paved the way for Brans and Dicke to formulate a very 
interesting theory of gravity \cite{BD} that is described by the 
metric tensor $g_{\mu\nu}$ and a scalar field $\Phi$. In their  
theory the Planck mass over an ensemble of universes is given 
by $M_{\rm p}^2(\Phi)= 16\pi\Phi$, 
and the Brans-Dicke (BD) coupling $\omega$ 
determines the validity of the principle of equivalence in 
gravitation. In the limit $\omega\to\infty$ and a suitably 
large $\Phi$, BD gravity is equivalent to General Relativity (GR). 
The theory predicts very small variations of $G$ within the horizon, 
but the value of $\omega$ is likely to leave its imprint on the 
CMB and in the different stages of the evolution of the FRW universe, 
such as the matter-radiation transition \cite{varyingG}. For larger 
scales, our implicit expectation is that although 
BD gravity is almost indistinguishable from GR in our observable 
universe, gravity may behave in a very different way in regions 
that are very distant from ours. 

Inflation has been on its own a very important tool 
in describing the early universe, and combined with 
BD gravity it enables us to envisage a very interesting quantum 
cosmological scenario where we can address the question of what  
are the {\it typical} values of the constants of 
Nature, either by means of anthropic arguments or by choosing  
suitable inflationary potentials. In chaotic inflation (for a 
review see e.g. \cite{inflation}) the scalar field that governs 
inflation starts out from large values and its classical 
``slow-roll'' motion along the slope of the potential 
towards the vacuum state is combined with quantum fluctuations. 
The fluctuations are stochastic \cite{stochastic} and they 
are responsible for continuously creating regions where inflation 
prevails, thus perpetuating the process indefinitely. 
The BD field influences the course of inflation, and the 
dynamical interplay of both scalar fields determines therefore 
the different stages of inflation, such as the beginning and 
end of inflation, the epoch when the classical drift and 
quantum fluctuations become comparable, etc. In extended inflation 
both fields evolve stochastically and the resulting distribution 
is to a large extent dependent on the potential.  
It can be argued that the assumption of starting out with 
large scalar fields in chaotic inflation in somewhat arbitrary. 
It can be shown however, that even though the physics may favour 
initial configurations where the fields are small (e.g. in the 
case of instanton solutions \cite{turok}), a Gaussian 
distribution of initial conditions around $\sigma\approx 0$ 
over a sufficiently broad ensemble of regions 
will result in an inflationary universe dominated by the 
largest values of $\sigma$ on the tails of this distribution, 
however strongly suppressed. 

Extended inflation has been investigated by a number of authors 
\cite{extended,extended2,extended3}, mostly in the context of first-order 
phase transitions of {\it old inflation}, where the so-called 
{\it big-bubble} problem arises as first pointed out by Liddle \& Wands 
\cite{bubble}. Later papers have focused on chaotic inflation, 
by computing the distributions of the fields, 
spectrum of density fluctuations, etc \cite{spectrum1,spectrum2},   
where transitions are second-order and there are no bubbles or 
discontinuous interfaces except for those created by quantum 
fluctuations. 

As it has been pointed out in \cite{V,mikel2}, the  
structure of the universe and values of the constants of 
Nature, as derived from the distributions of the fields,
depend on a very specific and crucial feature of the inflation 
potential. 
In the plane ($\sigma$,$\Phi$) it is easy to compute for each 
potential the end-of-inflation (EoI) and beginning-of-inflation 
(BoI) curves. The region of the plane enclosed between these two 
curves determines the range of values of the fields for which
inflation will take place. The classical trajectories span from 
BoI and cross EoI, but the quantum fluctuations allow jumps 
of the fields between neighbouring classical trajectories, and 
therefore the allowed states undergoing inflation quickly spread 
over the region enclosed between the curves BoI and EoI. The 
main characteristic of a potential is whether the area enclosed 
between BoI and EoI is finite or infinite. In \cite{V}, these 
are named {\it class I} and {\it class II} respectively. A 
{\it class II}  
potential yields to a universe where the spectrum of perturbations 
is arbitrarily small and the likelihood of a finite value of the 
Planck mass is negligible. In these theories, the values of the 
fields grow without limit, and any reasonable physical prediction 
becomes impossible without resorting to very stringent anthropic 
arguments. On the other hand, in the case of {\it class I} 
potentials BoI and EoI cross, and it is easy to show that the 
predominant values of the fields are those corresponding to the 
configuration exactly located at the crossing point. In this case, 
it is possible to predict values of the constants of Nature that 
are perfectly consistent with the observed values, and that comes 
out naturally from the physics, without an exaggerated use of 
anthropic arguments. In this paper we show that the exponential 
potential is a {\it class I} potential. 

In \S 2 we discuss the classical trajectories
of the fields in the slow-roll approximation and the form of
the BoI and EoI curves, which are the delimiters of inflation on the 
($\sigma$,$\Phi$) plane. In \S 3 we compute the probability
distributions $P(\sigma,\Phi)$, volume ratios of homogeneous 
hypersurfaces, and finally in \S 4 we derive the corresponding 
spectrum of density fluctuations discussing typical values of 
the parameters and consistency with the astrophysical constraints. 

\section{Classical trajectories}

The extended inflation action is given by \cite{extended} 
\begin{equation} 
\label{a}
S =\int d^{4}x \,\sqrt{-g}\left[\Phi R
   - {\omega\over\Phi}(\partial \Phi)^{2} 
 - \frac{1}{2}(\partial \sigma)^{2} \\
  - V(\sigma)\right] ,
\end{equation}
where $R$ is the curvature scalar and the potential is 
$V(\sigma) = V_0 e^{-\kappa\sigma}$. The coupling 
$\omega$ plays a similar r\^{o}le as that 
of the coupling functions $B_i(\Psi)$ of the dilaton field $\Psi$ 
in string theory. 
Based on this analogy, several authors have investigated 
the so-called {\it hyperextended inflation} models 
\cite{extended2,extended3,mikel2}, 
where $\omega$ becomes dependent on the BD field. 
In this paper however, we will merely examine the 
$\omega={\rm const}$ model. The BoI boundary is given by  
$V(\sigma)= M^{4}_{\rm p}(\Phi)$ or equivalently, 
\begin{equation}
\label{c1}
\Phi = \frac{V_0^{1/2}}{16\pi}\,e^{-\kappa\sigma/2}. 
\end{equation}
The BoI is the quantum limit where the metric 
fluctuations become significant and the inflaton field cannot 
take the values for which the potential is above this boundary.   
The EoI boundary is on the other hand 
\begin{equation}
\label{d}
\frac{1}{2}\dot \sigma^{2} + \omega\frac{\dot \Phi^{2}}{\Phi} 
\approx V(\sigma).   
\end{equation}
The equations of motion in a flat FRW background are 
\begin{equation}
\label{e1}
\left(D^2+\frac{1}{2\omega}R\right)\,\Phi=0\,, 
\end{equation}
\begin{equation}
\label{e2}
D^{2}\sigma = -V^{\prime}(\sigma)\,,
\end{equation}
\begin{equation}
\label{e3}
H^{2}+H\frac{\dot \Phi}{\Phi} = \frac{\omega}{6}
\left({\dot\Phi\over\Phi}\right)^2 + \frac{1}{6\Phi}
\left[\frac{1}{2}\dot\sigma^{2} + V(\sigma)\right] ,
\end{equation}
and the differential operator $D$ is defined 
\begin{equation}
\label{f}
D^{2} \equiv \partial^{2}_{t} + 3H\partial_{t} .
\end{equation}
In the slow-roll approximation, 
$\ddot \Phi \ll H\dot\Phi \ll H^{2}\Phi$ and 
$\dot \sigma^{2}+2\omega\,
\dot\Phi^{2}/\Phi 
\ll 2V(\sigma)$, (\ref{e1})-(\ref{e3}) read 
\begin{eqnarray}
\label{g1}
\frac{\dot\Phi}{\Phi} &=&2\frac{H}{\omega} \,, \\
\dot \sigma &=& -\frac{1}{3H}V^{\prime}(\sigma) \,,\\ 
H^{2} &=&\frac{1}{6\Phi}V ,
\end{eqnarray}
and the curvature scalar is given by $R=-12H^{2}$. The slow-roll
equations (\ref{g1}) enable us to rewrite (\ref{d}): 
\begin{equation}
\label{h}
\Phi_* =\left(3-\frac{2}{\omega}\right)\frac{1}{\kappa^{2}},
\end{equation}
where the $*$ subindex denotes the value at the end of inflation. 
Hence, $\Phi_*$ is independent of 
$\sigma$ and, for reasonably large $\omega$, it is solely determined 
by the slope of the potential. The condition $\Phi>0$ also imposes 
the constraint $\omega$, as can be seen in Fig.~1, such that 
the range $0<\omega<2/3$ is excluded to prevent imaginary values of the 
Planck mass. The classical trajectories of the fields are 
given by the following conservation law \cite{mikel2}
\begin{equation}
\label{h1}
\frac{d}{dt}\left[\omega\Phi + \int d\sigma 
\frac{V(\sigma)}{V^{\prime}(\sigma)}
\right]=0,
\end{equation}
which in the case of the exponential potential yields 
\begin{equation}
\label{h2}
\Phi= \frac{\sigma}{\kappa \omega} + \left(\Phi_0-
\frac{\sigma_0}{\kappa \omega}\right).
\end{equation}

In Fig.~2 we have put together the BoI and EoI curves and 
the classical trajectories of the fields on the ($\sigma$,$\Phi$) 
plane, i.e. (\ref{c1}),(\ref{h}) and (\ref{h2}) respectively. It can 
be seen in the figure that inflation takes place in the region
enclosed by BoI and EoI to the right of the intersection point 
$A$. The trajectories given by (\ref{h2}) are straight lines 
parallel to the segment $BC$, and the fields drift along these 
curves in the direction $B\to C$ during the course of inflation. 
The region enclosed by BoI and EoI to the left of $A$ does not 
undergo inflation, because the orientation of the classical 
trajectories is such that the fields would move from EoI towards 
BoI, which is not an acceptable solution. 
In addition to the classical trajectories quantum 
diffusion is responsible for the jumps of the fields between
neighbouring classical trajectories. It can be seen that, unlike 
with powerlaw potentials, for which $\sigma$ decreases as $\Phi$ 
increases during the course of inflation, in the case of the 
exponential potential both fields increase during the slow roll. 

\begin{figure}[t]
\centering
\leavevmode\epsfysize=5.5cm \epsfbox{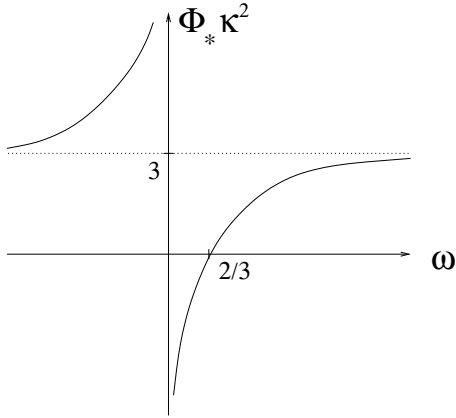}\\
\vskip 0.2cm
\caption[fig1]{BD field at EoI, $\Phi_*$, 
vs. $\omega$ for an arbitrary value of $\kappa$. 
$\Phi_*$ is given in units of $\kappa^{-2}$. 
Inflation takes place in the range $\omega<0$ or $\omega\geq 2/3$.}
\end{figure}

\begin{figure}[t]
\centering
\leavevmode\epsfysize=5.5cm \epsfbox{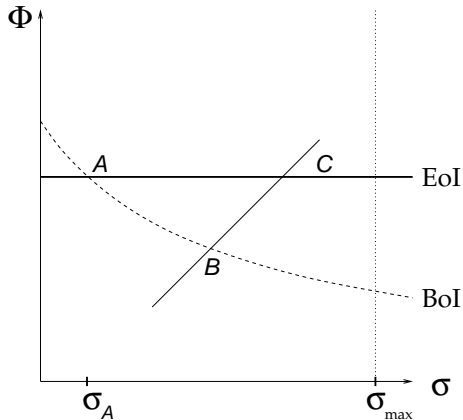}\\
\vskip 0.2cm
\caption[fig2]{Predicted BoI and EoI curves, 
dashed and thick solid curves respectively. Classical 
trajectories are straight lines parallel to $BC$. At the intersection 
point $A$ the onset and end of inflation coincide. $\sigma_{\rm max}$ 
determines the scale of validity of the slow-roll
approximation. Inflation takes place within the region enclosed by
BoI, EoI and $\sigma\approx\sigma_{\rm max}$.}
\end{figure}

The EoI boundary (\ref{h}) gives a definite and 
unique prediction for $\Phi_*$, and also it implies that 
if $\Phi_0>\Phi_*$ inflation will not occur. In the case of 
$0<\Phi_0\lsim\Phi_*$, inflation takes place for values of the inflaton 
\begin{equation}
\label{sigma0}
\sigma_0 \gsim -\frac{2}{\kappa}\, {\rm log}\left({16\pi\Phi_0
\over V_0^{1/2}}\right).
\end{equation} 
Naturally if $\Phi_0=\Phi_*$, then the RHS of (\ref{sigma0}) 
is $\sigma_A$, the value of the field at the intersection point 
$A$ of BoI and EoI in Fig.~2. 

It must be noted that the slow-roll approximation does not 
hold in the case of the exponential potential for arbitrarily 
large values of $\sigma$. For a given value of $\kappa$ it is 
straightforward to compute $\sigma_{\rm max}$ for which the 
potential and kinetic energy of the fields are comparable and 
thus the slow-roll conditions break down. It is easy to show from 
(\ref{d}) that this scale is 
\begin{equation}
\label{smax}
\sigma_{\rm max}\approx \left({3\omega-2\over\kappa}\right) .
\end{equation}
Therefore it follows that the EoI boundary does not span 
from $\sigma_A$ to infinity, but inflation will occur within 
a finite region $\sigma_A\lsim\sigma\lsim\sigma_{\rm max}$.


\section{Likelihood ratios}

In the framework of extended inflation we can predict the 
distribution of $G$ in an ensemble of universes, depending on 
the scalar potential (see e.g. \cite{extended2,extended3,spectrum1}). 
If the 
likeliest values predicted differ to a large extent from those 
observed in our region of the universe, then it becomes necessary 
to invoke the anthropic principle to justify the theory. On the 
other hand, a more optimal situation is achieved if the potential 
is able to yield maxima in the distribution of the fields that 
are compatible with the observed data. We have shown in the 
previous section that in the case of an exponential potential 
the outcome at the end of inflation is simple, and regardless 
of the initial conditions, a period of inflation leads to a uniform 
distribution of $\Phi$. As a result, regions that are still 
inflating will have values of the BD field in the range 
$0<\Phi<\Phi_*$, those regions where inflation has ended have 
invariably $\Phi=\Phi_*$. There will be other regions whose 
initial conditions do not yield inflation (i.e. they are not 
located in the region enclosed by BoI and EoI in Fig.~2). 
The dominant contribution to the total volume will be given 
by the regions that are still inflating or have completed 
inflation in the recent past. 

It is apparent from Fig.~2 that whereas the magnitude of $\Phi$ 
is bounded during the course of inflation, that of the inflaton 
scalar is not. We can thus expect that quantum fluctuations 
may prolong the course of inflation by taking $\sigma$ to arbitrarily 
large values. In this section we compute the volume ratios 
of homogeneous hypersurfaces ($\sigma$,$\Phi$)= const with 
respect to the total volume occupied by thermalized regions, 
using the results of \cite{mikel2}. As 
inflation is eternal, the volumes of the hypersurfaces are divergent, 
though the volume ratios are not. These ratios give a good measure 
of the relative likelihood of an arbitrary configuration ($\sigma$,$\Phi$) 
with respect to another. 

The comoving probability $P_{c}(\sigma,\Phi,t)$ is governed 
by the conservation equation \cite{stochastic,stationary}
\begin{equation}
\label{continuity}
\partial_{t}P_{c} =-\partial_{\sigma}J_{\sigma} 
- \left({\Phi\over 2\omega}\right)^{1/2}\!\! \partial_{\Phi}
J_{\Phi},
\end{equation}
where the probability current $\vec{J} \equiv (J_{\sigma},J_{\Phi})$ is 
given by the slow-roll solutions in the regime where quantum 
diffusion is neglected 
\begin{eqnarray}
\label{jsigma}
J_{\sigma} \approx -\frac{M_{\rm p}^{2}(\Phi)}{4\pi} H^{-1} 
\partial_{\sigma}H P_{c} \,, \\
\label{jphi}
J_{\Phi} \approx -\frac{M_{\rm p}^{2}(\Phi)}{2\pi} 
\left({\Phi\over 2\omega}\right)^{1/2}\!\!
H^{-1} \partial_{\Phi}H P_{c}, 
\end{eqnarray}
where $t$ corresponds to cosmic time. In order to solve 
(\ref{continuity}), it is customary to adopt the eigenvalue 
expansion \cite{extended3,stationary}
\begin{equation}
\label{eigen}
P_{c}(\sigma,\Phi,t) = \sum_{n=1}^{\infty}\psi_{n}(\sigma,\Phi) \\
e^{-\gamma_{n}t} ,
\end{equation}
where $\gamma_{1} < \gamma_{2} < \gamma_{3} < ...,$ and thus the 
asymptotic limit $t\rightarrow \infty$ yields 
$P_{c} \sim \psi_{1} e^{-\gamma_{1}t}$. Substituting 
(\ref{jsigma})(\ref{jphi}) in (\ref{continuity}), we get \cite{mikel2}
\begin{equation}
\label{exp-prob}
P_{c}(\sigma,\Phi,t) \approx C_{0}\Phi^{1/2}
\, \exp(8\pi\omega\gamma_1\Phi-\gamma_1 t) ,
\end{equation}
where $C_{0}$ is a normalization constant and $\gamma_1$ is 
estimated numerically from the resulting eigenvalue equation for 
different values of $V_0$ and $\kappa$. For $V_0\sim 1$ and 
$0.1M_{\rm p}\lsim \kappa^{-1}\lsim 10M_{\rm p}$, we have 
$2.4\lsim \gamma_1\lsim 2.8$.  

The volume ${\cal V}_{*}$ of thermalized regions over the entire 
spacetime is determined by the two-dimensional probability flux of
the fields across the EoI boundary (\ref{h}). The line element along 
EoI is $d\sigma$ and the differential flux is 
$d\sigma ( \vec{J}\cdot \hat{n})$, where $\hat{n}$ is a normal 
vector to EoI. Hence 
\begin{equation}
\label{vol*}
{\cal V}_{*} = {\cal V}_{0}\left|\int_{0}^{t_{c}} dt \,e^{3t} 
\int_{\sigma_A}^{\sigma_{\rm max}} d\sigma
J_{\Phi}\right|,
\end{equation}
where ${\cal V}_0$ is the initial homogeneous volume, 
$t_c$ is a cutoff time that we use to regularize the volumes 
of the hypersurfaces, following the method employed in 
\cite{method,method2,mikel2}. Therefore, substituting 
(\ref{exp-prob}) in (\ref{jphi}), we get 
\begin{equation}
\label{vol*2}
{\cal V}_{*} = {\cal V}_{0}\left(\frac{4C_0}{\kappa}\right)
\,{\Phi_*\over(2\omega)^{1/2}}\, 
(\sigma_{\rm max}-\sigma_A)
e^{8\pi\omega\gamma_1\Phi_*}\,
\frac{e^{(3-\gamma_{1})t_{c}}}{ 3-\gamma_{1}}.
\end{equation}
On the other hand, the volume ${\cal V}(\sigma,\Phi)$ of 
arbitrary hypersurfaces $(\sigma,\Phi)$=const that are undergoing 
inflation is determined by the probability flux at 
($\sigma$,$\Phi$) across a line element located along the classical 
trajectory at that point. I.e., 
$|(\vec{J}\cdot\hat{l})dt|$, where $\hat{l}$ is the tangent
vector to (\ref{h2}). The expression is  
\begin{equation}
\label{volume}
{\cal V}(\sigma,\Phi) = {\cal V}_{0}\left|\int_{0}^{t_{c}}
dt \,e^{3t} (\vec{J}
\cdot\hat{l})\right|,
\end{equation}
and thus
\[
{\cal V}(\sigma,\Phi) = {\cal V}_0{2C_0\kappa\over
(1+\omega^2\kappa^2)^{1/2}}\,\Phi\,
\left[\kappa\omega\Phi^{1/2}+{1\over\kappa}\left(\frac{2}{\omega}
\right)^{1/2}\right]
\]
\begin{equation}
\label{volume2}
\times \exp(8\pi\omega\gamma_1\Phi)\,
\frac{e^{(3-\gamma_{1})t_{c}}}{3-\gamma_{1}}.
\end{equation}
Therefore the volume ratio $r$ of the hypersurface 
$(\sigma,\Phi)$ with respect to the thermalized regions is given by
\[
r =\frac{{\cal V}(\sigma,\Phi)}{{\cal V}_{*}} 
\propto \Phi\,
\left[\kappa\omega\Phi^{1/2}+{1\over\kappa}\left(\frac{2}{\omega}
\right)^{1/2}\right]
\]
\begin{equation}
\label{ratio}
\times\exp\big[8\pi\omega\gamma_1(\Phi-\Phi_*)\big] 
\sim \Phi^{3/2}\,\exp(24\pi\omega\Phi),
\end{equation}
which is totally independent of $\sigma$, as are 
(\ref{exp-prob})(\ref{volume2}). We note the tendency 
towards larger values of $\Phi$, and indeed the likeliest value 
that is attained is $\Phi_*$. The ensembles of hypersurfaces 
$\Phi=$const that are undergoing inflation are equally likely 
and occupy the same fraction of the total volume. These correspond to the 
horizontal cross-sections in the range $0<\Phi<\Phi_*$ in Fig. 2. 
In the limit of large $\omega$ this trend is preserved, except in 
the case when $\omega <0$. In that case, (\ref{ratio}) yields the 
greatest likelihood for the smallest values of $\Phi$, i.e. 
$\Phi\approx\Phi_0$. 

\section{Spectrum of fluctuations}

In order to calculate the resulting spectrum of fluctuations 
we need to derive the amplitudes of the typical quantum fluctuations 
in this model. In addition to the classical drift given by the
slow-roll solutions, the stochastic nature of inflation exerts a 
quantum force over distances larger than $H^{-1}$ that can be 
described by \cite{stochastic}
\begin{eqnarray}
\label{sdot}
\dot\sigma &=& 
- \frac{1}{3H}V^{\prime}(\sigma)+\frac{H^{3/2}}{2\pi}\zeta(t) \,\\
\label{pdot}
\dot\Phi &=& 2\frac{H}{\omega}\Phi + 
\frac{H^{3/2}}{2\pi}\left(\frac{\Phi}{2\omega}\right)^{3/4}\!\xi(t),
\end{eqnarray}
where the Gaussian variables $\zeta$,$\xi$ satisfy   
$\langle \zeta(t_{1})\zeta(t_{2})\rangle 
= \langle\xi(t_{1})\xi(t_{2})\rangle = \delta(t_{1}-t_{2})$
and $\langle\zeta(t_{1})\xi(t_{2})\rangle =0$. Naturally (\ref{sdot}) 
applies in the slow-roll regime, such that $\sigma_A\lsim\sigma
\lsim\sigma_{\rm max}$, and for arbitrarily large fields 
$\sigma\gg\sigma_{\rm max}$ the classical kinetic and potential 
energies are negligible, and $\sigma$ becomes stationary, whereas 
quantum fluctuations lead to changes in $\Phi$. The quantum 
jumps $(\delta\sigma,\delta\Phi)$ around an arbitrary hypersurface 
($\sigma$,$\Phi$) are distributed according to 
\begin{equation}
\label{jumps}
dP(\delta\sigma,\delta\Phi)\sim \frac{{\cal V}(\Phi +\delta\Phi)}
{{\cal V}(\Phi)} \,dP_{0}(\delta\sigma,\delta\Phi) ,
\end{equation}
where $dP_{0}$ is a Gaussian distribution 
\begin{equation}
\label{gaussian}
dP_{0}(\delta\sigma,\delta\Phi) \propto  
\exp\left[-{(\delta\sigma)^{2}\over 2\Delta_1^2} 
-{(\delta\Phi)^{2}\over 2\Delta_2^{2}}\right]
d\delta\sigma d\delta\Phi,
\end{equation}
where $\Delta_1\equiv H/2\pi$ and $\Delta_2\equiv 
(\Phi/2\omega)^{1/2} H/2\pi$ are the variances of the fields 
and the preceding factor in the 
RHS of (\ref{jumps}) is the volume ratio that determines the relative 
likelihood of the configuration $\Phi+\delta\Phi$ with respect to 
$\Phi$. From (\ref{ratio}) we have 
\begin{equation}
\label{factor}
\frac{{\cal V}(\Phi +\delta\Phi)}
{{\cal V}(\Phi)} \approx \left(1 +{\delta\Phi\over\Phi}
\right)^{3/2}(1+8\pi\omega\gamma_1\,\delta\Phi),
\end{equation}
and therefore the typical quantum jumps, given by the stationary 
values of (\ref{jumps}), are 
\begin{equation}
\label{sjump}
\langle\delta\sigma\rangle \approx 0,
\end{equation}
\begin{equation}
\label{pjump}
\langle\delta\Phi\rangle \approx 
(\gamma_1\Phi)^{1/2}\,{H\over 4\pi},
\end{equation}
where the fluctuations in $\sigma$ remain Gaussian whereas the 
fluctuations of the BD field are not, due to the presence of the 
factor (\ref{factor}). As we have discussed in the previous section, 
in the case of $\omega>2/3$, the largest values of $\Phi$ are the 
likeliest ones, reaching a maximum at $\Phi\approx\Phi_*$. From 
(\ref{pjump}) it follows that in this case the typical jumps are 
positive, and the evolution of the field towards $\Phi\approx\Phi_*$ 
is enhanced by quantum fluctuations of order $\sim\Phi^{1/2}H$. 

In the case of $\omega<0$, we have seen from (\ref{ratio}), 
that the smallest values of $\Phi$ are enhanced, i.e. 
$\Phi\approx \Phi_0$, 
and second-order corrections to (\ref{pjump}) contribute to 
suppress quantum jumps towards larger values of $\Phi$. The 
amplitude of these corrections is 
$\sim \omega V/\Phi^2$, 
and it is only significant for small values of $\sigma$ and $\kappa$ and 
large $\omega$. As one departs from $\Phi\approx\Phi_0$ 
towards larger values of $\Phi$, the leading order of the 
fluctuations (\ref{pjump}) rapidly dominates. 

In the Einstein frame, the adiabatic density fluctuations with 
$\langle\delta\sigma\rangle\approx 0$ over 
distance scales of $H^{-1}$ are given by 
\begin{equation}
\label{Hwave}
\frac{\delta \rho}{\rho} = -\frac{12}{5} H\omega\,
\frac{\dot\Phi}{\Phi}\,\left(\dot\sigma^2 +2\omega
\frac{\dot\Phi^2}{\Phi}\right)^{-1}\delta\Phi,
\end{equation}
and therefore, by substituting (\ref{pjump}) we get 
\begin{equation}
\label{spectrum}
\biggl\langle{\delta\rho\over\rho}\biggr\rangle
\approx \frac{1}{10\pi}{H\over\Phi^{1/2}},
\end{equation}
which is to be evaluated for $N=65$ $e$-foldings after 
crossing EoI. The value of $\Phi$ in (\ref{spectrum})  
is roughly $\Phi\approx \Phi_*$, as it will not change 
significantly after inflation, and $H$ is evaluated 
at $\sigma\approx \sigma_{\rm max}$. If the theory is 
correct, then the uniformity of the distribution (\ref{h}) 
at EoI implies that $M_{\rm p}^{*}\sim 10^{19}$ GeV throughout, 
as in our region of the universe, and 
$\kappa\sim 10^{-18}~~{\rm GeV}^{-1}$, and from 
(\ref{smax}) we have that $\sigma_{\rm max}\sim 10^{21}$ 
GeV. From the astrophysical constraint  
$\langle\delta\rho/\rho\rangle \lsim 10^{-4}$, these estimates 
in turn imply that 
$V_0\lsim 10^{58}\,e^{6\omega}~~{\rm GeV}^4,$ 
which in practical terms leaves $V_0$ unconstrained for 
conservative values of $\omega\gsim 500$.

\section{Conclusions}

In this paper we have examined extended inflation with an exponential 
potential. 
The remarkable feature of this model is 
the prediction of a constant distribution of the 
Planck mass at the end of inflation, given by (\ref{h}). 
The parameter $\kappa$ of the theory 
is therefore estimated via the observed Planck mass in this region of 
the universe, which in turn fixes the parameter $\sigma_{\rm max}$ 
that determines the range of values of $\sigma$ for which inflation 
takes place. 

The amplitude of the potential $V_0$ is left unconstrained by 
astrophysical bounds on the spectrum of fluctuations, as described 
by the argument given in \S 4. The dynamics as is given in \S 2 and 
the likelihood distributions in \S 3 are shown to be insensitive 
to the numerical value of this parameter. 

As is shown in Fig.~2, the BoI and EoI curves in this model 
cross at $\sigma=\sigma_A$ and the area enclosed between them is 
thus infinite. However the breakdown of the slow-roll approximation 
for the exponential potential over the range 
$\sigma\gsim \sigma_{\rm max}$ (where $\sigma_{\rm max}$ is given 
by (\ref{smax})) implies that in practical terms only a finite 
region of the ($\sigma$,$\Phi$) plane undergoes inflation. In 
the classification of \cite{V} this means that the exponential 
potential is {\it class I}, i.e. the values of the fields at 
EoI remain finite.


\subsection*{Acknowledgments}

A.M. is supported by the Inlaks foundation and an ORS award.  
We thank Kei-ichi Maeda, Andrew Liddle, John Barrow, 
Dominik Schwarz and Reza Tavakol for useful discussions,
and A.M. acknowledges use of the Starlink computer system at the University
of Sussex.

\end{document}